\documentclass[twocolumn,showpacs,amsmath,amssymb,prl]{revtex4}

\usepackage{epsfig}
\usepackage{color}
\usepackage{times}
\usepackage{pst-all}

\begin{document}

\title{Scale Invariance and Nonlinear Patterns of Human Activity}

\author{Kun Hu$^{1}$, Plamen~Ch.~Ivanov$^{1}$$^{2}$, Zhi Chen$^{1}$,
Michael~F.~Hilton$^{3}$, H.~Eugene~Stanley$^{1}$, Steven~A.~Shea$^{3}$
}

\affiliation{$^{1}$Center for Polymer Studies and Department of Physics, Boston University, Boston, MA 02215\\
$^{2}$Harvard Medical School, Beth Israel Deaconess Medical Center, Boston, MA 02115\\
$^{3}$Harvard Medical School and Division of Sleep Medicine, Brigham \&
Women's Hospital,Boston, MA 02115
}

\begin{abstract}
 We investigate if known extrinsic and intrinsic factors fully account for
 the complex features observed in recordings of human activity as
measured from forearm motion in subjects undergoing their regular daily
 routine. We demonstrate that the apparently random
 forearm motion possesses previously
 unrecognized dynamic patterns characterized by fractal and nonlinear
 dynamics. These patterns are unaffected by changes in the average activity
 level, and persist  when the same subjects undergo time-isolation laboratory
 experiments  designed to account for the circadian phase and to
 control the known  extrinsic factors. We attribute these patterns to a novel
 intrinsic multi-scale dynamic regulation of human activity.
\end{abstract}

\received{March 17, 2003}
\revised{August 01, 2003}
{Code number: LC9266} 
%\date{\today}  

\maketitle
Control of human
activity is complex, being influenced by many factors both
extrinsic (work, recreation, reactions to unforeseen random
events) and intrinsic (the circadian pacemaker that influences our
sleep/wake
cycle~\cite{ForcedDesy_Czeisler_Science1999} and ultradian oscillators with shorter time
scales~\cite{ultradian_Kleitman_Sleep}). The extrinsic factors may account
for the apparently random fluctuations in human motion observed
over short time scales while the intrinsic rhythms may account for the underlying regularity
in average activity level over longer periods of up to 24 h.
Further, human activity correlates with important physiological
functions including whole body oxygen consumption and heart
rate.

\begin{figure}
\centerline{\epsfysize=8.8cm{{\epsfbox{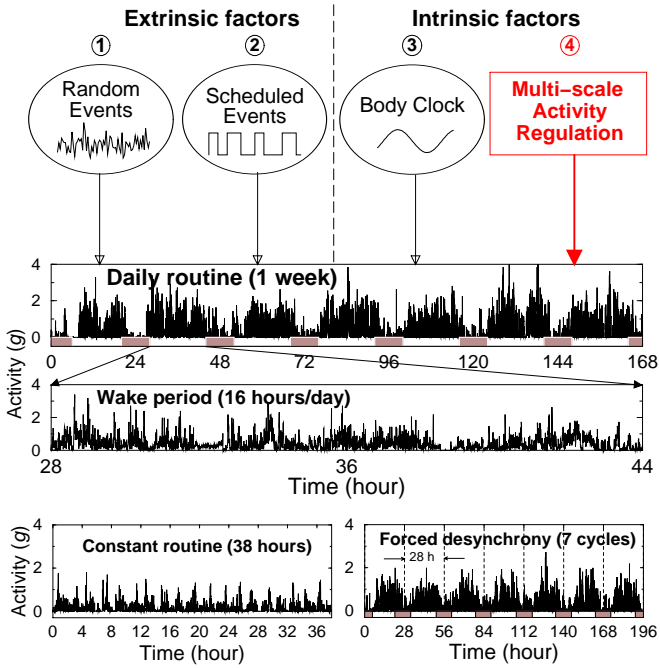}}}}
%\centerline{\epsfysize=8.8cm{\rotate[r]{\epsfbox{fig1.eps}}}}
\caption{\label{fig1} 
  Independent contributors to the complex dynmaics of human activity,
  depicted at the top of the figure, include: \textcircled{1} reaction to
  extrinsic random events, \textcircled{2} scheduled activities and,
  \textcircled{3} intrinsic factors, notably the endogenous circadian
  pacemaker which influences the sleep/wake cycle.  Our findings of
  scale-invariant activity patterns (Figs.~\ref{fig2},~\ref{fig3}) indicate a heretofore-unidentified
  intrinsic multi-scale control of human activity
  ~\textcircled{\textcolor{red}{4}}, which is independent of other extrinsic
  and intrinsic factors such as \textcircled{1}, \textcircled{2}, and
  \textcircled{3}.  The second panel illustrates an actual one-week
  recording of human activity~\cite{actiwatch} during the daily routine
  protocol. Data structure highlights a 24-h sleep/wake periodic change in
  the mean activity --- lowest during sleep (filled bars).  The third panel, expanding a 16-h
  section of wakefulness, also shows patches of high and low average activity
  levels with apparent erratic fluctuations at various time scales. The bottom left panel is an
  activity recording from the same subject during the constant routine
  protocol with much lower average activity values compared to daily routine. The clear 2-h cycle is a result of scheduled
  laboratory events.  The bottom right panel shows activity levels in the
  same subject during the forced desynchrony protocol.  Of note here is the
  28-h sleep/wake cycle as opposed to the 24-h rhythm in activity data during
  the daily routine.}
\end{figure}

{\it Actiwatch} devices~\cite{actiwatch} are traditionally used to
demarcate sleep versus wakefulness based on average activity
levels, or to observe the mean pattern of activity as it changes
across the day and night according to disease state
(Fig.~1).  Traditionally 
 activity fluctuations are considered as random noise and have been ignored. We hypothesize that there are systematic patterns in
the activity fluctuations that may be independent of known extrinsic and
intrinsic factors. 

To test our hypotheses, we evaluate the structure of human activity during
wakefulness, using: (i) probability distribution analysis; (ii) power
spectrum analysis, and (iii) fractal scaling and nonlinear analysis.  To
elucidate the presence of an intrinsic activity control center
independent of known circadian, ultradian, scheduled and random factors, we
apply 3 complementary protocols.

$\bullet$ (A) {\it Daily routine protocol}: We record activity data
throughout two consecutive weeks in 16 healthy ambulatory domiciliary
subjects (8 males, 8 females, 19-44 years, mean 27 years) performing their
routine daily activities. The only imposed constraints are that subjects go
to bed and arise at the same time each day (8 h sleep opportunity) and that
they are not permitted to have daytime naps (Fig.~\ref{fig1}).

$\bullet$ (B) {\it Constant routine protocol}: To assess intrinsic activity
controllers (i.e. circadian or other neural centers) independent of scheduled
and random external influences activity recordings are made in the laboratory
throughout 38 h of constant posture (semi-recumbent), wakefulness,
environment ($21^{o}C$, dim light [$< 8$ lux]), dietary intake and scheduled events~\cite{ConstR_Czeisler_Science1989_ConstR_Brown_JBiolRhy1992}.  This protocol is performed in a subset of subjects (7 males, 4 females)
that participated in the daily routine protocol. These highly controlled and
constant experimental conditions result in reduced average and variance of
activity levels.

$\bullet$ (C) {\it Forced desynchrony protocol}: To test for the presence of
heretofore unidentified intrinsic activity control centers, independent of
known activity regulators (circadian pacemaker), whilst accounting for
scheduled and random external influences, we employ the validated ¡Forced
desynchrony¢ (FD) protocol~\cite{ForcedDesy_Czeisler_Science1999}.  Six (4 male, 2 female) of the 16 subjects
that participated in the daily routine protocol completed the FD limb of the
study. For eight days subjects remain in constant dim light (to avoid ``resetting'' the body clock). Sleep periods are delayed by 4 h every day,
such that subjects live on recurring 28 h ``days'', while all scheduled
activities become desynchronized from the endogenous circadian pacemaker.
Thus, as measurements occur across all phases of the circadian clock, the
effect of intrinsic circadian influences can be removed~\cite{ForcedDesy_Czeisler_Science1999}.  Average
activity level and activity variance are also significantly reduced due to
laboratory-imposed restrictions on the subject¢s activity (Fig.~\ref{fig1}).

%To test our hypotheses, we perform three sets of complementary analyses of
%the data during wakefulness: (i) probability distribution analysis; (ii) power spectrum
%analysis, and (iii) fractal scaling and nonlinear analysis. 

\begin{figure}
\centerline{\epsfxsize=8.5cm{\epsfbox{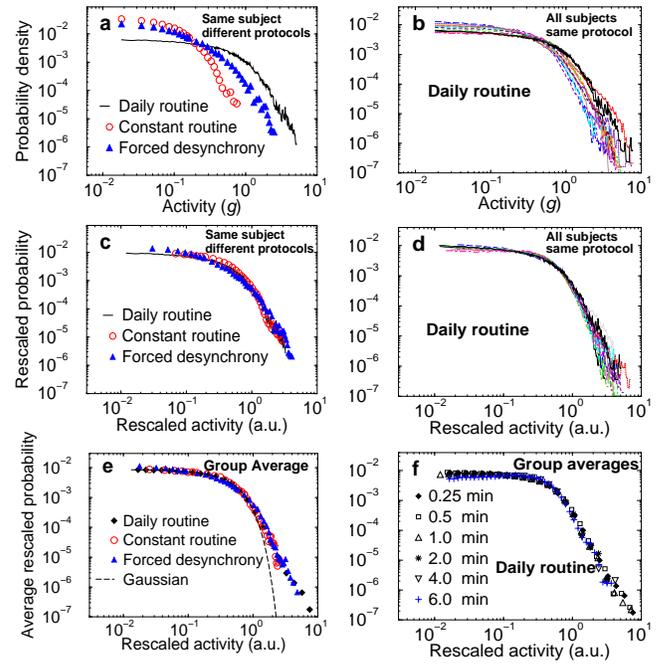}}}
\caption{\label{fig2}Probability distributions of activity values
during wakefulness. {\bf (a)} Probability distributions for an
individual subject during 14 consecutive days of daily routine, 38
h of constant routine and 8 days of the forced desynchrony
protocol.  {\bf (b)} Probability distributions for 16 subjects
during the daily routine protocol indicate large difference
between individuals.
{\bf (c)} Same probability distributions as in (a), after
appropriately rescaling both axes. 
Data points for all three protocols collapse onto a single curve.
{\bf (d)} Same probability distributions as in (b), after
rescaling. {\bf (e)} Group average of the rescaled distributions
during all three protocols.
Again all distributions collapse onto a single curve.
{\bf (f)} Group
average of all individual distributions rescaled as in (e) obtained for
varied time windows  during the daily
routine.}
\end{figure}

When the same subject is studied in
different protocols, we find large differences in the probability
distributions (Fig.~2). For example, during wakefulness greater values of activity
occur most frequently during the daily routine, intermediate activity values occur
during the forced desynchrony, and the highest frequency of low activity
values is seen during the constant routine (Fig.~2a). Indeed, the largest
activity values encountered during the constant routine protocol are
approximately two orders of magnitude less frequent than similar activity
values encountered in the daily routine protocol. We also find major
differences between individuals in the distribution of activity values during
the daily routine protocol (Fig.~2b). Such differences are expected given their different daily schedules,
environments, and reactions to random events.  However, by appropriately
rescaling the distributions of activity values on both axes to account for
differences in average activity level and standard deviation~\cite{rescaling},
we find a remarkable similarity in the shapes of the probability
distributions for all three protocols (Figs. 2c, 2e), and for all individuals
when in the same protocol (Fig.~2d). The existence of a {\it universal} form of
the probability distribution, independent of activity level in all
individuals and in all protocols, suggests that a {\it common} underlying
mechanism may account for the overall distribution of activity.

This probability distribution when plotted on a log-log
scale reveals different characteristics above and below a distinct
crossover point (Fig.~2e). At scales above the crossover activity
level there is pronounced non-Gaussian tail (Fig.~2e). This tail
on the log-log plot represents a power-law form, indicating an
intrinsic self-similar structure for a range of activity values.
Moreover, we find that the observed shape of the rescaled
probability distribution remains unchanged when the data series
are reanalyzed using a variety of observation windows ranging from
15 s to 6 min (Fig.~2f). This stability of the probability
distribution over a range of time scales indicates that the
underlying dynamic mechanisms controlling the activity have
similar statistical properties on different time scales.
Statistical self-similarity is a defining characteristic of
fractal objects and is
reminiscent of a wide class of physical systems with universal
scaling properties.
Our finding of a universal form of the probability distribution raises the possibility of an intrinsic mechanism
that influences activity values in a self-similar ``fractal''
manner, that is unrelated to the individual's daily and weekly
schedules, reactions to the environment, the average level of
activity, the phase of the circadian pacemaker, and the time scale
of observation.

\begin{figure}
\centerline{ \epsfxsize=7.6cm{{\epsfbox{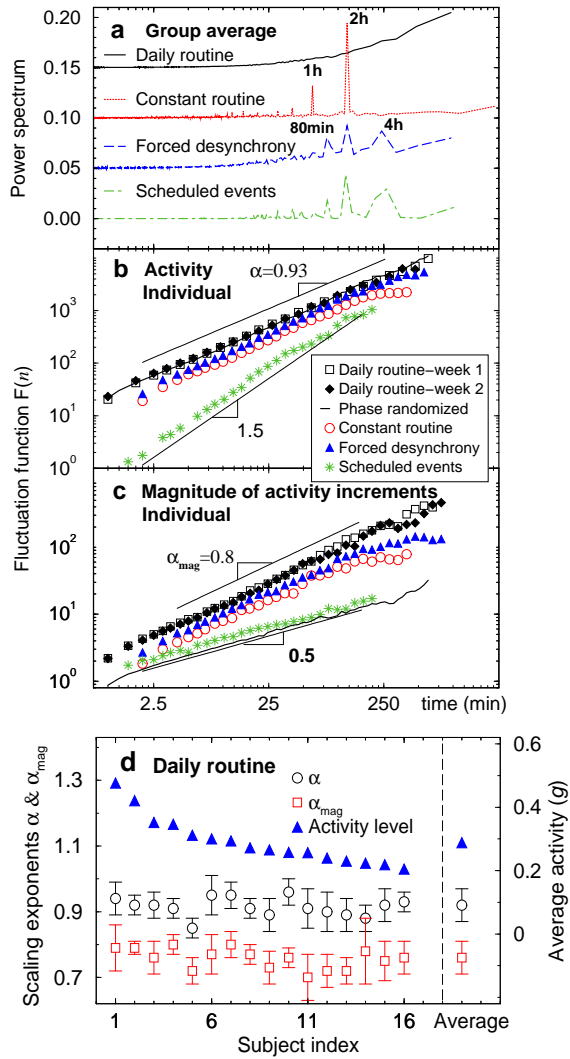}}} }
\caption{\label{fig3}Power spectrum and correlation analyses of activity data
during wakefulness. {\bf (a)} Group average power spectral densities for all three protocols. For
better clarity and to avoid overlap, curves are vertically offset.  To present graphs on a common x-axis, power spectra are shown with decreasing frequency from left to right.  The spectral density
peaks for the simulated scheduled activity data representing controlled
scheduled events during the protocol (bottom curve) match the peaks observed in the
original human activity data recorded during the forced desynchrony
protocol. This analysis suggests that the observed peaks in the power
spectrum cannot be attributed to endogenous ultradian rhythms.
{\bf (b)} DFA scaling of activity for an individual during wakefulness
throughout two separate weeks of daily routine, $38$ h of constant routine,
and 8 days of forced desynchrony protocols. {\bf (c)} DFA scaling of
the magnitude series of activity increments for the same signals as in (a).
A scaling exponent $\alpha_{\rm mag} \approx 0.8 $ of similar value is observed for all
three protocols.  {\bf (d)}
Scaling exponents $\alpha$ and $\alpha_{\rm mag}$ (left scale), and average activity
levels (right scale) for all 16 subjects obtained from a 14-day daily routine
protocol. Although the average activity level between subjects changes considerably (from
$0.2$ to $0.5$), both scaling exponents are consistent for all subjects, exhibiting
a group average of $\alpha=0.92\pm 0.05$ and $\alpha_{\rm mag}=0.77\pm0.05$.
}
\end{figure}

We next perform power spectral analyses for all three protocols to determine
whether there exist any systematic intrinsic ultradian rhythms of
activity with periods of less than 24 h
duration~\cite{ultradian_Kleitman_Sleep,monkey}.
The data for each individual exhibit occasional peaks in the
daily routine protocol for periods ranging from 30 min to 4 h. However, we
find no systematic ultradian rhythms within individuals from week to week,
and no systematic ultradian rhythms in the group average for the daily
routine protocol (Fig.~3a). The only systematic rhythms that are ostensibly
in the ultradian range which emerge in the group data are at 4 h during the
forced desynchrony protocol (with harmonics at 2 h and 80 min) and at
2 h during the constant routine protocol (with harmonics at 1 h and 30 min)
(Fig.~1 and Fig.~3a). These peaks are caused by the controlled scheduled activities in
the laboratory and are extrinsic to the body as they also occur in simulated
scheduled activity
data that assumes specific activity values for each scheduled behavior
imposed throughout the laboratory protocols (Fig.~3a). Thus, we find no
evidence of systematic intrinsic ultradian rhythms in our data.

To provide further insight into the dynamic control of
activity, we next examine the temporal organization in the
fluctuations in activity values.  We perform detrended fluctuation analysis
(DFA) %~\cite{dfa_Bul} 
which quantifies correlations in the activity
fluctuations after accounting for nonstationarity in the data by
subtracting underlying polynomial trends. The DFA method
quantifies the root mean square fluctuations, $F(n)$, of a signal
at different time scales $n$. Power-law functional form,
$F(n)\sim n^{\alpha}$, indicates self-similarity (fractal
scaling). The parameter $\alpha$, called the scaling exponent,
quantifies the correlation properties in the signal: if
$\alpha$=0.5, there is no correlation (random noise); if $\alpha<0.5$, the signal is
anticorrelated, where large activity values are more likely to be
followed by small activity values; if $\alpha>0.5$, there are
positive correlations, where large activity values are more likely
to be followed by large activity values (and vice versa for small
activity values).

Figure 3b shows that $F(n)$ for a typical subject during
wakefulness exhibits a power-law form over time scales from
$\approx 1$ min to $\approx 4$ h. We find that the scaling
exponent $\alpha$ is virtually identical for records obtained
during the first week of daily routine ($\alpha = 0.92 \pm 0.04$, mean $\pm$
standard deviation among subjects), the second week ($\alpha=0.92
\pm 0.06$) of the daily routine, the constant routine protocol
($\alpha = 0.88 \pm 0.05$), and the forced desynchrony protocol
($\alpha = 0.92 \pm 0.03$). The value of $\alpha \approx 0.9$ for
all protocols and all individuals indicates that activity
fluctuations are characterized by strong long-range positive
correlations. Furthermore, we find that this scaling behavior is not caused by
the scheduled activities because simulated scheduled activity data that are generated
by assigning a specific activity value for each scheduled event
throughout the laboratory protocols yields an exponent of $\alpha
= 1.5$ (Fig.~3b), which represents random-walk type
behavior. %~\cite{shlesinger_west}. 
These results suggest that the activity fluctuations are not a
consequence of random events (in which case $\alpha$ would be
0.5) or scheduled events, but rather relate to an underlying
mechanism of activity control with stable fractal-like features
over a wide range of time scales from minutes to hours. Since mean
activity levels and the amplitude of the fluctuations are greatly
reduced  in the laboratory during the constant routine and forced
desynchrony protocols (Fig.~1), we obtain smaller values of $F(n)$
(downward shift of the lines in Fig.~3b). However there is no
change in the scaling exponent $\alpha$. Similarly, the scaling
exponents for the daily routine protocol are independent of the
average activity levels of the different subjects (Fig.~3d), and
of the mean activity level on different days of the week, indicating that this newly-found scaling pattern of activity appears to be an intrinsic feature.

To test for the presence of nonlinear properties of the
data, we analyze the ``magnitude series'' formed by taking the
absolute values of the increments between consecutive activity
values{~\cite{mag_sign}}.  Again, from detrended fluctuation analysis of
this series, we find practically identical
scaling exponents, $\alpha_{\rm mag}$, for all three protocols,
despite large differences in mean activity levels between
protocols (Fig~3c). Moreover, all individuals have very similar values of
the scaling exponent $\alpha_{\rm mag}$ (Fig.~3d), which are not
systematically changed by the protocol. For the group, during the
first week of daily routine, we find $\alpha_{\rm mag} = 0.78 \pm
0.06$ (mean $\pm$ standard deviation among subjects), during the
second week $\alpha_{\rm mag} = 0.76 \pm 0.05$, during the
constant routine protocol $\alpha_{\rm mag} = 0.82 \pm 0.05$, and
during the forced desynchrony protocol $\alpha_{\rm mag} = 0.80
\pm 0.04$. Since $\alpha_{\rm mag}\approx 0.8 (>0.5)$, there are
positive long-range correlations in the magnitude series of
activity increments, indicating the existence of nonlinear
properties related to Fourier phase interactions (Fig.~3c)~\cite{mag_sign,multifractal99}. To confirm that the
observed positive correlations in the magnitude series indeed
represent nonlinear features in the activity data, we do the
following test: we generate a surrogate time series by performing
a Fourier transform on the activity recording from the same
subject during daily routine as in Fig.3b, preserving the amplitudes of the Fourier
transform but randomizing the phases, and then performing an
inverse Fourier transform. This procedure eliminates
nonlinearities, preserving only the linear features of the
original activity recording such as the power spectrum and correlations. Thus, the new surrogate signal has the same
scaling behavior with $\alpha = 0.93$ (Fig.~3b) as the original activity recording;
however, it exhibits uncorrelated behavior for the magnitude
series ($\alpha_{\rm mag} = 0.5$) (Fig.~3c). Our results show that
the human activity data contains important phase correlations
which are canceled in the surrogate signal by the randomization
of the Fourier phases, and that these correlations do not exist in
the simulated scheduled activity. Further, our tests indicate
that these nonlinear features are related to Fourier phase
interactions, suggesting an intrinsic {\it nonlinear}
mechanism~\cite{multifractal99}. The similar value of $\alpha_{\rm
mag}$ for all three protocols and all individuals, which is
different from $\alpha_{mag} = 0.5$ obtained for the simulated
scheduled activity and for the phase randomized data, confirms
that the intrinsic dynamics possess nonlinear features that are
independent of the daily and weekly schedules, reaction to the
environment, the average level of activity, and the phase of the
circadian pacemaker.

The consistency of our results among individuals, and
for different protocols, suggests that there exist previously unrecognized
complex dynamic patterns of
human activity that are unrelated to extrinsic factors or to the
average level of activity. We also showed these patterns to be
independent of known intrinsic factors related to the circadian
and to any ultradian rhythms. Notably, (i) these patterns are
unchanged when obtained at different phases of the circadian
pacemaker; (ii) we do not observe systematic intrinsic ultradian
rhythms in activity among subjects in the daily routine
experiment; (iii) imposing strong extrinsic ultradian rhythms at 4
h and 2 h in the laboratory protocols did not change the fractal
scaling exponents $\alpha$ or $\alpha_{mag}$ or the form of the
probability distribution; and (iv) we find consistent results over
a wide range of time scales. Together, these findings strongly
suggest that our results are unlikely to be a reflection of the
basic rest activity cycles or ultradian rhythms.
We attribute these novel
scale-invariant patterns to a robust {\it intrinsic multi-scale}
mechanism of regulation (Fig.~1). This regulatory
mechanism presents a new challenge to
understand nonlinear control of human motor activity and 
pathways of interaction with other physiologic dynamics such as heart
rate,  gait~\cite{gait_Haudsdorff_gait_Golubitsky_Nature}, finger
tapping~\cite{human_coordination_Chen_YQ}, human sway and muscle
fluctuations~\cite{humansway_Chow_HMS}.


\begin{thebibliography}{99}
\bibitem{rescaling} 
To test if the individual probability density curves follow a common
functional form, we rescale the curves while preserving the normalization to
unit area. We divide the activity values by a constant, $A_0$, and multiply the
probability density function by the same constant, where $A_0$ is the activity
value before rescalingof each individual curve for which the cumulative
probability (i.e., the area under the density function curve) is $60\%$.
\end{thebibliography}
\end{document}